\documentclass[twocolumn,scriptaddress,secnumarabic,amssymb,nobibnotes, aps,prd]{revtex4-1}

\newcommand{\revtex}{REV\TeX\ }
\newcommand{\classoption}[1]{\texttt{#1}}
\newcommand{\macro}[1]{\texttt{\textbackslash#1}}
\newcommand{\m}[1]{\macro{#1}}
\newcommand{\env}[1]{\texttt{#1}}
\usepackage{tikz}
\usepackage{array}
\usepackage{units}
\usepackage{amsmath}
\usepackage{multirow}
\usetikzlibrary{positioning} 
\usepackage{tabularx}
\usepackage{hhline,float}
\newfloat{Tabelle}{thp}{lot}
\usepackage{slashbox}
\usepackage{subfigure}
\usepackage{nicefrac}
\usepackage{textcomp}
\usepackage{esvect}
\usepackage{booktabs}
\usepackage{verbatim}
\usepackage{graphicx}
\usepackage{caption}
\usepackage{epstopdf}
\usepackage{braket}
\usepackage[hidelinks]{hyperref}
\hypersetup{pdfcreator=N/A}
\hypersetup{pdfproducer=N/A}
\hypersetup{pdfauthor=N/A}
\hypersetup{pdfsubject=N/A}

\begin{document}

\title{Closed clusters approach to graphene}%

\author{Ilja I. Taljanskij}%
\email[Electronic address: ]{taljanilja@gmail.com}
\affiliation{Department for Theoretical Physics, Ivan Franko National University of Lviv}
\altaffiliation{Current address: Rudolf-Breitscheid-Stra{\ss}e 39, 23968 Wismar, Germany}
\date{\today}%

\begin{abstract}

 The Closed Cluster method (CC method) is applied to find solutions for various calculation problems of the energy band structure of graphene. The essence of the CC method consists in the addition of closing bonds between edge atoms to the usual cluster method in order to eliminate the "dangling" bonds on the edges of the cluster. We study the cases of an "infinite" layer of graphene as well as nanoribbons, nanotubes and bilayer graphene. Results for these cases are in agreement to that what was obtained by means of other methods (tight binding approximation and others). By means of the CC method we also study the problem of point defects in graphene and obtain the distortion of the energy spectrum. The energy spectrum of the layer C$_{1-x}$ Si$_{x}$ $(0\leq x \leq 1)$ is found as well as the dependence of the energy gap on the concentration of silicon. We show that the energy band structure of C$_{1-x}$ Si$_{x}$ looks like a tunnel transition. Wave functions of graphene in the symmetry points of Brillouin zone are also obtained.

\end{abstract}

\maketitle

\tableofcontents

\section{Introduction}
\label{sec:intro}
Studies of various properties of graphene and it`s applications have attracted much attention in the last years, as it is well-known \cite{Wakabayashi1}. 
In this paper, it is proposed for an approach based on the use of closed clusters (CC) to calculate the energy band structure of graphene. We have developed this approach earlier in application to the one-dimensional and three-dimensional crystals with diamond structures \cite{Tal1}. The essence of the CC approach is to bring together all bonds of atoms which are located on the edge of a cluster in order to eliminate the ``dangling'' bonds. This approach is found to be especially useful for calculation of the energy spectrum of crystals with point defects, such as vacancies or impurity atoms. The simplest and most widespread approach to calculate such impurity states is known to be the effective mass method. This approximation works sufficiently well in cases of impurity levels being located along the borders of the energy zones. However, it is inapplicable to the description of so-called ``deep levels'', which lie far from the zone borders \cite{Lannoo}. One of the methods used for those levels is the cluster approach, in which a group of atoms - the cluster - is mentally picked from a crystal lattice. 

This cluster is considered as a separate ``molecule'' and for it`s calculation the usual quantum chemistry methods are applied. The advantage of the clusters approach is it`s applicability for modelling real situations of impurity atoms and - if necessary - taking into consideration a possible distortion of the crystal lattice. 

However, the usual cluster approach has a deficiency. Is the ``infinity'' crystal substituted by a group of atoms, a problem with atoms lying on the edge of the cluster arises. The presence of such atoms with torn bonds distorts the energy spectrum of the crystal. This distortion can be diminished by increasing the size of the cluster, but the approximation to the exact value is very slow.

This deficiency can instead be removed by connecting the torn bonds with each other and hereby closing them. A similar procedure, the so-called ``periodical boundary conditions'', is applied in studying the energy spectrum of the infinite ideal crystals. The special feature of our approach is the application the closing procedure of these bonds to small clusters to study crystals with distorted regularity properties (such as impurity, edges and other).
In this paper, the CC approach is applied to graphene, a relatively new material with numerous of unique properties \cite{Geim}. In section\,\ref{sec:essence} the fundamental idea of the CC approach is presented as well as the rules for the construction of diagrams corresponding to the various clusters. In section\,\ref{sec:Ham} examples of building Hamiltonian matrices and solutions for the secular equations in case of periodical structures without edges and defects are given. The aim of this section is to test our approach by means of comparison of the results with those obtained by other methods. In section\,\ref{sec:Bilayer graphene} the CC approach is applied to bilayer graphene. In section\,\ref{sec:Wave functions} we present wave functions at symmetry points of the Brillouin zone. In section\,\ref{sec:Nanoribbons and nanotubes} we study nanoribbons and nanotubes by means of closed clusters. Section\,\ref{sec:Impurity} contains the calculations of the graphene energy spectrum in the presence of impurity. Finally, in \ref{sec:monoatomic layer} we study the energy band structure of the hypothetic monoatomic layer C$_{1-x}$ Si$_{x}$ $ \; (0\leq x \leq 1)$. Section\,\ref{sec:Con} concludes with a discussion of the obtained results and furthermore describes a possibility for making the approach more precise as well as the application of the CC approach to other problems. 

\section{The essence of the CC approach and a rule for constructing the diagrams}
\label{sec:essence}

By the term ``closed cluster'' we mean a group of N atoms which reflects the structure of the crystal and furthermore the atom bonds which are present within the group - at the same time the torn bonds are absent. A similar procedure of closing the bonds known as ``periodical boundary conditions' is often used in one-dimensional atom chains when atom N+1 is identical to the first one. Such a procedure is also possible for 3D-systems, but is rarely applied here, since closing a 3D cluster is much more difficult. In 2D cases however, closing can be realized very easily - therefore, application of the CC approach to graphene and other 2D structures seems to be very effective.

\begin{figure}[hptb!]
\centering
\subfigure[\label{Fig1a}]{\includegraphics[width=2.9cm]{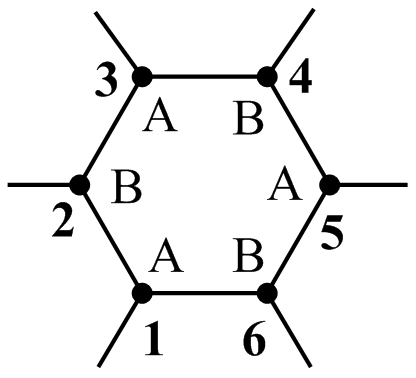}}\qquad
\centering
\subfigure[\label{Fig1b}]{\includegraphics[width=2.2cm]{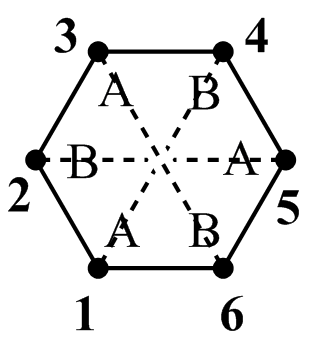}}\qquad
\subfigure[\label{Fig1c}]{\includegraphics[width=3.7cm]{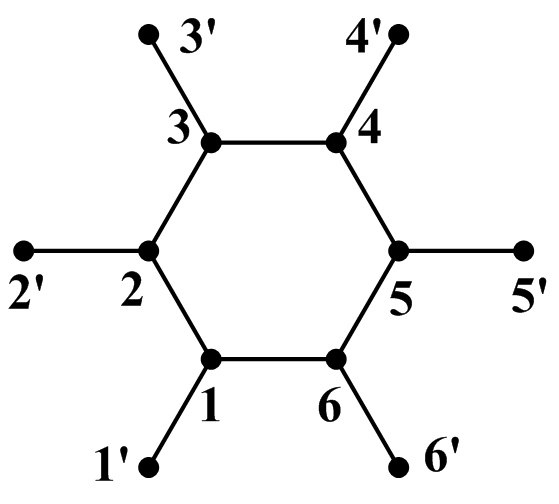}}
\caption{The most simple clusters of the graphene lattice\\(a) unclosed cluster $N=6$ \\ (b) closed cluster $N=6$ \\ (c) cluster surrounded by neighbor atoms \\
dashed line: closing bonds \\ continuous line: real bonds in graphene lattice}
\label{Fig1}
\end{figure}

While constructing clusters corresponding to graphene one must start out from it`s crystal structure. The latter is well-known, a hexagonal layer which is formed by two sublattices, here A and B \cite{Wallace}. The simplest unclosed cluster corresponding to such a structure is shown in Fig.\,\ref{Fig1a}.

The main idea of the closed cluster approach is the necessity of closing the torn bonds shown in Fig.\,\ref{Fig1a} in the way shown in Fig.\,\ref{Fig1b}. In case of an endless graphene layer the structure of the lattice requires the following rules regarding the construction of a closed cluster with approximation of the nearest neighbors.

\begin{comment}
\begin{figure}[htbp]
\centering
\includegraphics{./1}
\caption{ The most simple clusters of the graphene lattice\\(a) unclosed cluster $N=6$ \\ (b) closed cluster $N=6$ \\ (c) cluster surrounded by neighbor atoms \\
dashed line: closing bonds \\ continuous line: real bonds in graphene lattice}
\label{Fig. 1}
\end{figure}
\end{comment}

\begin{itemize}
\item \textbf{\textit{Rule 1.}} Each atom of sublattice A is bonded with three atoms of sublattice B and vice versa.

\item \textbf{\textit{Rule 2.}} All real and closing bonds have the same energy levels. 

\item \textbf{\textit{Rule 3.}} The number of atoms in a cluster N must be divisible by six.
\end{itemize}

The last rule is explained in detail in section\,\ref{sec:Wave functions}. 

All three rules in fact are satisfied within the construction in Fig.\,\ref{Fig1b}, although the second rule seems to be broken at first sight: for example, the closing bond between atoms 1 and 4 looks different from that between 1-2 or 1-6, since atom 4 is further away from 1. However, it must be kept in mind that the bond 1-4 is not a real bond. In distinction from Fig.\,\ref{Fig1c}, the closed cluster 1(b) must be understood as a diagram or graph, which makes it easier to obtain the Hamiltonian matrix. Furthermore, if these Hamiltonian matrix elements corresponding to the closing bonds are chosen the same as for existing bonds, then the ``interaction'' between atoms like 1 and 4 in 1(b) in fact describes the interaction between 1 and 1' in 1(c), latter of which is absent in cluster $N=6$.

It must be noted, that for the numbers of atoms in sublattices A and B it is necessary to be equal, as closing bonds is only possible between atoms of different sublattices. For example, in cluster 1(b) it is only possible to close the bonds of atom 1 with atom 4, but not with 3 or 5. This restriction is necessary to satisfy Rule 1.

We also note that energy values obtained from the solution of secular equations are independent from the choice of the cluster in case of an ``infinite'' ideal lattice, but are only dependent from the number N. They are as well independent from the numeration of the atoms, since changes in numeration only cause determinant permutations. 

\begin{figure}[hptb!]
\centering
\subfigure[\label{Fig2a}]{\includegraphics[width=4.7cm]{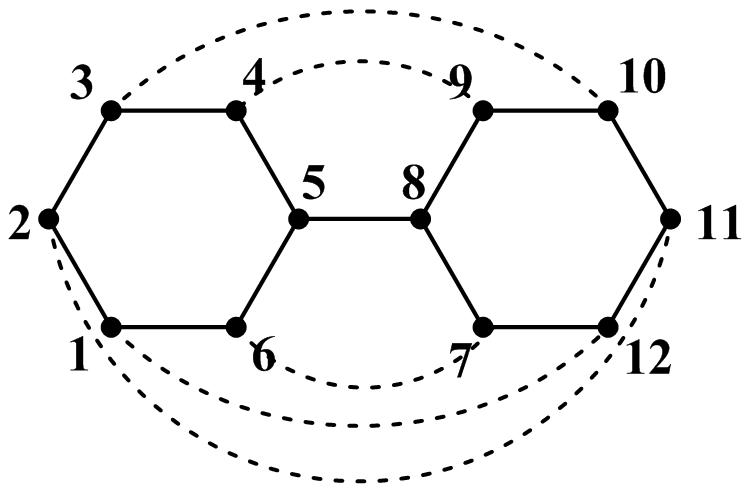}}\qquad
\centering
\subfigure[\label{Fig2b}]{\includegraphics[width=3cm]{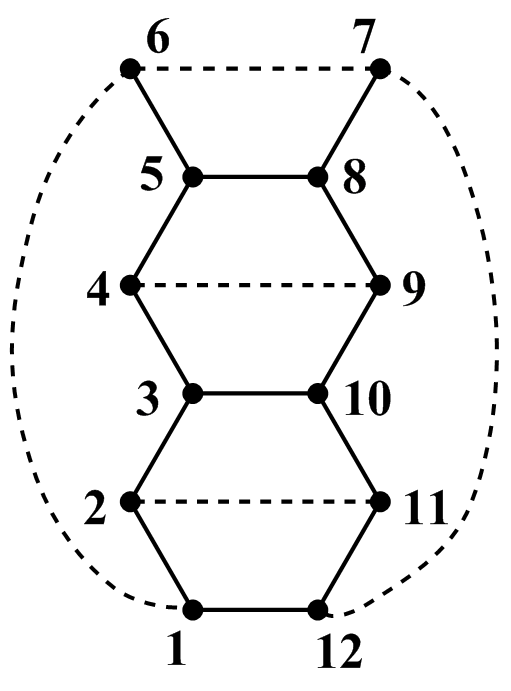}}\qquad
\subfigure[\label{Fig2c}]{\includegraphics[width=3.7cm]{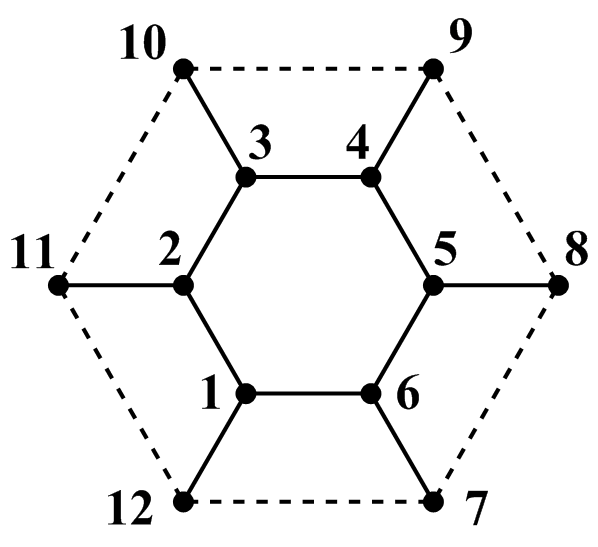}}
\caption{Examples of closed clusters for $N=12$.}
\label{Fig2}
\end{figure}

Fig.\,\ref{Fig2} shows some of the possible clusters with $N=12$ with dissimilar ways of closing.

%\begin{tikzpicture}[color=black]
%\draw (60:1cm) -- (120:1cm) -- (180:1cm) -- (240:1cm) -- (300:1cm)
%-- (360:1cm) -- cycle;
%\end{tikzpicture}

%\begin{tikzpicture}
%[inner sep=0.5mm,
%place/.style={circle, draw=black!50,fill=black!80},]
%\node at ( 0.5,1) [place] {};
%\coordinate [label=left:\textcolor{black}{$4$}] (1) at (0.5,1);
%\draw [--]
%\node at ( 1,0) [place] {};
%\coordinate [label=left:\textcolor{black}{$5$}] (1) at (1,0);
%\node at ( 0.5,-1) [place] {};
%\coordinate [label=left:\textcolor{black}{$6$}] (1) at (0.5,-1);
%\node at ( -0.5,-1) [place] {};
%\coordinate [label=left:\textcolor{black}{$1$}] (1) at (-0.5,-1);
%\node at (-1,0) [place] {};
%\coordinate [label=left:\textcolor{black}{$2$}] (1) at (-1,0);
%\node at (-0.5,1) [place] {};
%\coordinate [label=left:\textcolor{black}{$3$}] (1) at (-0.5,1);
%\end{tikzpicture}

To conclude this section, we note, that the name "closed cluster" which we use is different from the term "closed walk" used in graph theory \cite{Trinajstic}. For example, the cluster shown in Fig.\,\ref{Fig1a} is a "closed walk", but not a closed cluster - the latter is shown in Fig.\,\ref{Fig1b}. 

One definition of a closed cluster can be given as a cluster, where each atom is linked to the same number of neighbor atoms as in the corresponding crystal. 

\section{Building of a Hamiltonian matrix and solution of secular equations}
\label{sec:Ham}

The basic idea underlying the CC approach is the same as in the usual molecular orbital approach (MO) \cite{Heilbronner}. In particular, this is the representation of the wave function of the cluster $\Psi (\underline{r})$ as a linear combination of the wave functions of the atoms
\begin{equation} \label{eq:1}
\Psi (\textbf{r}) = \sum\limits_{n=1}^{N} c_n \varphi  (\bold{r} - {\bold a_n}).
\end{equation}

Functions $\varphi$ in graphene are $\ket{p_z}$ - orbitals of carbon atoms, with axis z being perpendicular to the layer. We designate wave functions along and opposed this axis as "+" and "-".

The standard procedure of obtaining secular equations leads to a system of linear equations which help to find the coefficients $c_n$ in
\begin{equation} \label{eq:2}
\sum\limits_{n=1}^{N} M_{mn} c_n =  0 , \qquad  m= 1,2,\ldots , N ,
\end{equation}

where
\begin{equation} \label{eq:3}
M_{mn} = \varepsilon \delta_{mn} + (1-\delta_{mn}) \eta_{mn},
\end{equation}

$\delta_{mn} = \begin{cases} 1, m=n \\   0, m \neq n, \end{cases}$

$\eta_{mn} = \begin{cases} 1, \textnormal{if atom m is bound with atom n}  \\  0, \textnormal{if atom m is not bound with atom n,} \end{cases}  $

\begin{equation*}
\varepsilon = \frac{E-E_0} {\gamma_0}  
\end{equation*}
with

$E$ - energy of an electron

$E_0$ - energy of an electron in the $\ket{2p_z}$ state in an isolated carbon atom. We use $E_0=0$

and

$-\gamma_0$ - the transfer integral between neighbor atoms with wave functions with the same sign, $\gamma_0 > 0$.

The problem of calculating the corresponding values of energy $\varepsilon$ (in $\gamma_0$ units) is then reduced to solving the secular equation 
\begin{equation} \label{eq:4}
\det {M_{mn} = 0}.
\end{equation}

Now we consider some examples of applications of the approach. We begin with the simplest closed cluster $N=6$, which is shown in Fig.\,\ref{Fig1b}. The secular equation (4) has the form (empty matrix cells standing here and further for zeroes):
\\[1em]

\begin{equation} \label{eq:5}
 D_{6} =	\left|   
 \begin{array}  {rrrrrr} 
\varepsilon & 1 & & 1 & & 1  \\
1 & \varepsilon & 1 & & 1 &  \\
& 1 & \varepsilon & 1 & & 1 \\
1 & & 1 & \varepsilon & 1 &  \\
& 1 & & 1 & \varepsilon & 1 \\
1 & & 1 & & 1 & \varepsilon  \\
\end{array}
\right| = 0.
\end{equation}

Calculating the determinant \eqref{eq:5} leads to the following equation for finding $\varepsilon$:
\begin{equation} \label{eq:6}
\varepsilon^4 (\varepsilon^2 - 9) = 0.
\end{equation}

The solutions of \eqref{eq:6} $\pm$3,0,0,0,0 are exactly those energy values at bottom, top and Dirac points of two energy bands, which are obtained from a tight binding and nearest neighbor approximations [5].

The fact, that the CC approach provides exact values of energy band boundaries already at a minimal cluster size is very important in the calculation of the energy of impurity states, as the energy of such states are counted from the bands' boundaries.

Let us consider the next example with $N=12$. If we proceed from the clusters from Fig.\,\ref{Fig2}, then the secular equation has the form
\\[1em]
\newline
\small

\begin{equation} \label{eq:7}
 D_{12} =	\left|   
 \begin{array}  {rrrrrrrrrrrrr} 

 	 \varepsilon & 1 &   &   &   & 1 &   &   &   &  &  & 1\\
 	 1 & \varepsilon & 1 &  &  &  &  &  &  &  & 1 &\\
 	  & 1 & \varepsilon & 1 &  &  &  &  &  & 1 &  &\\
 	  &  & 1 & \varepsilon & 1 &  &  &  & 1 &  &  &\\
 	  &  &  & 1 & \varepsilon & 1 &  & 1 &  &  &  &\\
 	 1 &  &  &  & 1 & \varepsilon & 1 &  &  &  &  &\\
 	  &  &  &  &  & 1 & \varepsilon & 1 &  &  &  &1\\
 	  &  &  &  & 1 &  & 1 & \varepsilon & 1 &  &  &\\
 	  &  &  & 1 &  &  &  & 1 & \varepsilon & 1&  &\\
 	  &  & 1 &  &  &  &  &  & 1 & \varepsilon& 1&\\
 	  & 1 &  &  &  &  &  &  &  & 1 & \varepsilon& 1 \\
 	  1&  &  &  &  &  & 1 &  &  &  & 1 &\varepsilon \\
 
 	 \end{array}\right|
=0.
\end{equation}
\\[2em]
\newline
The solution of equation \eqref{eq:7} is given by following values of energy $\varepsilon$: $\pm 3, \pm 2, \pm 2, \pm 1, 0, 0, 0, 0$.

The determinant of cluster N=24 has a form which is analogous to \eqref{eq:7} $D_{12}$. The obtained energy values are shown in Table\,\ref{Table1}.

\begin{table*}[htbp]
\scriptsize
\caption{Electron energy values with varying numbers of atoms in the cluster }
\begin{center}
\begin{tabular}{c|c|c|c|c|c|c|c|c|c|c|c|c|c|c}
\hline
{N} & \multicolumn{ 14}{c}{\tiny{Energy}} \\ \hline
{6} & {$\pm$ 3} &  &  &  &  &  &  &  &  &  & {0} & {0} & {0} & {0} \\
{12} & {$\pm$ 3} &  &  & {$\pm$ 2} & {$\pm$ 2} &  & {$\pm$ 1} &  &  &  & {0} & {0} & {0} & {0} \\
{24} & {$\pm$} 3 & {$\pm (\sqrt{3}$+1)}  & {$\pm (\sqrt{3}$+1)}  & {$\pm$ 2} & {$\pm$ 2} & {$\pm$ 1} & {$\pm$ 1} & {$\pm$ 1} & {$\pm (\sqrt{3}$-1)}  & {$\pm (\sqrt{3}$-1)}  & {0} & {0} & {0} & {0} \\ \hline
\end{tabular}
\end{center}
\label{Table1}
\end{table*}

The main conclusion which may be drawn from the comparison of cases $N=6$, $12$ and $24$ is the fact, that the energy values of the bottoms and tops of the lower $(-3,0)$ and upper band $(0,3)$ are equal in each case, therefore they are independent of the cluster size. With the growth of the cluster size new energy levels arise, however, previous levels remain the same.

Each energy value in Tab.\,\ref{Table1} is corresponding with some point of the Brillouin zone. More in-detail discussion on that issue is presented in sec.\,\ref{sec:Wave functions}At this point we just briefly discuss the fourfold degeneration of level $\varepsilon=0$ at all $N$. 

\begin{figure}[htbp!]
\centering
\includegraphics[width=3.5cm]{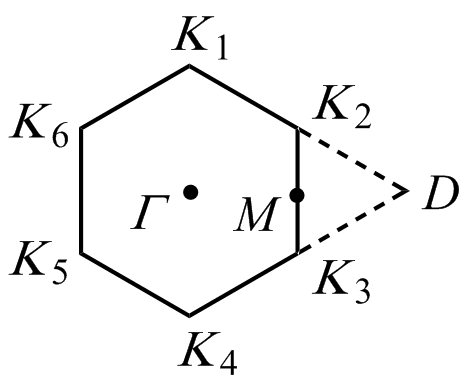}
\caption{Brillouin zone of graphene \cite{Wallace}.}
\label{Fig3}
\end{figure}

At first glance it seems not to be in accordance with $\varepsilon$ being 0 at six corners of the Brillouin zone (points $K_1$, $K_2$, \ldots , $K_6$ in Fig.\,\ref{Fig3}).

However, it must be kept in mind, that only two of these six points, $K_1$ and $K_2$ for example, belong to the first Brillouin zone, others belonging to the next zones. The twofold degeneration at points $K_1$ and $K_2$ is what leads to fourfold degeneration of the level $\varepsilon=0$.

In conclusion of this section it should be noticed, that the rules for diagram construction formulated in sec.\,\ref{sec:Ham} does not require these diagrams to be plane. Therefore, they also can be applied to nanotubes as well as spherical surfaces (fullerene). Stratified cluster can also be used. For example, the diagram in Fig.\,\ref{Fig2a} can be rearranged to a hexahedron without breaking the bonds. The application for the space diagram is useful for studying many-layer graphene. In the next section we use a three-dimensional diagram to study bilayer graphene.

\section{Bilayer graphene}
\label{sec:Bilayer graphene}

\begin{figure}[htbp!]
\centering
\includegraphics[width=5.5cm]{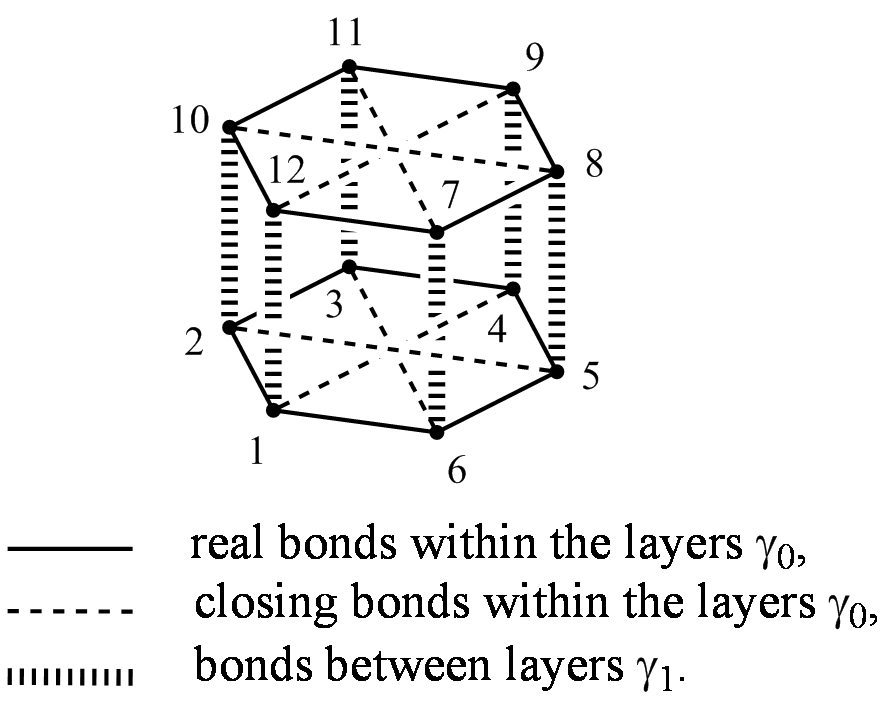}
\caption{Cluster $N=12$ for bilayer graphene.}
\label{Fig4}
\end{figure}

For $N=12$, the 3D closed cluster corresponding to bilayer graphene is shown in Fig.\,\ref{Fig4}. In addition to the closing bonds within the layers (energy $\gamma_0$), also the bonds describing the interactions between those layers (energy $\gamma_1$ in $\gamma_0$-units) are present in the cluster.

Note, that Fig.\,\ref{Fig4} does not show the displacement of the upper layer towards the lower, since this Figure doesn´t reproduce a real atomic structure, but is solely a diagram which is used for building a Hamiltonian matrix.

The secular equation corresponding to Fig.\,\ref{Fig4} has the form

\small
\begin{equation} \label{eq:8}
 D_{12}^{(2)} =	\left|   
 \begin{array}  {rrrrrrrrrrrrr} 

 	 \varepsilon & 1 &   & 1  &   & 1 &   &   &   &  &  & \gamma_1\\
 	 1 & \varepsilon & 1 &  & 1 &  &  &  &  &  & \gamma_1 &\\
 	  & 1 & \varepsilon & 1 &  & 1 &  &  &  & \gamma_1 &  &\\
 	 1 &  & 1 & \varepsilon & 1 &  &  &  & \gamma_1 &  &  &\\
 	  & 1 &  & 1 & \varepsilon & 1 &  & \gamma_1 &  &  &  &\\
 	 1 &  & 1 &  & 1 & \varepsilon & \gamma_1 &  &  &  &  &\\
 	  &  &  &  &  & \gamma_1 & \varepsilon & 1 &  & 1 &  &1\\
 	  &  &  &  & \gamma_1 &  & 1 & \varepsilon & 1 &  & 1 &\\
 	  &  &  & \gamma_1 &  &  &  & 1 & \varepsilon & 1&  &1\\
 	  &  & \gamma_1 &  &  &  & 1 &  & 1 & \varepsilon& 1&\\
 	  & \gamma_1 &  &  &  &  &  & 1 &  & 1 & \varepsilon& 1 \\
 	  \gamma_1&  &  &  &  &  & 1 &  & 1 &  & 1 &\varepsilon \\
 
 	 \end{array}\right|=0.
\end{equation}

Solving equation \eqref{eq:8} leads to following energy levels:

\begin{equation} \label{eq:9}
\varepsilon = \pm (3+\gamma_1), \pm (3-\gamma_1), \pm \gamma_1, \pm \gamma_1, \pm \gamma_1, \pm \gamma_1.
\end{equation}

The main result that ensues from \eqref{eq:9} is that there is an energy gap $\varepsilon_g = 2\gamma$. The presence of this gap in bilayer graphene is a fact discussed in many papers \cite{Castro}\cite{Seyller}. In the model which we are using, the energy gap is caused by the interaction between the layers and disappears if we set $\gamma_1$=0.

\section{Wave functions}
\label{sec:Wave functions}

Within our considered model the cluster wave functions are determined by the totality of coefficients $c_n$, according to \eqref{eq:1}. To obtain the latters one must solve the system of equations \eqref{eq:2} under defined energy values $\varepsilon$. For $\varepsilon = 3$ and $\varepsilon = -3$ the result is obvious and shown in Tab.\,\ref{table2}. Let us now consider the cases $\varepsilon = \pm $1 and $\varepsilon$ = 0 in more detail. Latter is most interesting, since at this point the valence band meets the conduction band.

\begin{comment}
$$
\begin{array}  {ccccccccccccc} 

 	(1)& \varepsilon & 1 &   & 1  &   & 1 &   &   &   &  &  & \gamma_1\\
 	 (2)&1 & \varepsilon & 1 &  & 1 &  &  &  &  &  & \gamma_1 &\\
 	 (3)& & 1 & \varepsilon & 1 &  & 1 &  &  &  & \gamma_1 &  &\\
 	  (4)&&  & 1 & \varepsilon & 1 &  &  &  & \gamma_1 &  &  &\\
 	  (5)&& 1 &  & 1 & \varepsilon & 1 &  & \gamma_1 &  &  &  &\\
 	 (6)&1 &  & 1 &  & 1 & \varepsilon & \gamma_1 &  &  &  &  &\\
 	  (7)&&  &  &  &  & \gamma_1 & \varepsilon & 1 &  & 1 &  &1\\
 	  (8)&&  &  &  & \gamma_1 &  & 1 & \varepsilon & 1 &  & 1 &\\
 	  (9)&&  &  & \gamma_1 &  &  &  & 1 & \varepsilon & 1&  &1\\
 	  (10)&&  & \gamma_1 &  &  &  & 1 &  & 1 & \varepsilon& 1&\\
 	  (11)&& \gamma_1 &  &  &  &  &  & 1 &  & 1 & \varepsilon& 1 \\
 	  (12)&\gamma_1&  &  &  &  &  & 1 &  & 1 &  & 1 &\varepsilon \\
 \label{D12b}
 	 \end{array}
$$
\end{comment}

\begin{figure}[htbp]
\centering
\includegraphics[width=6cm]{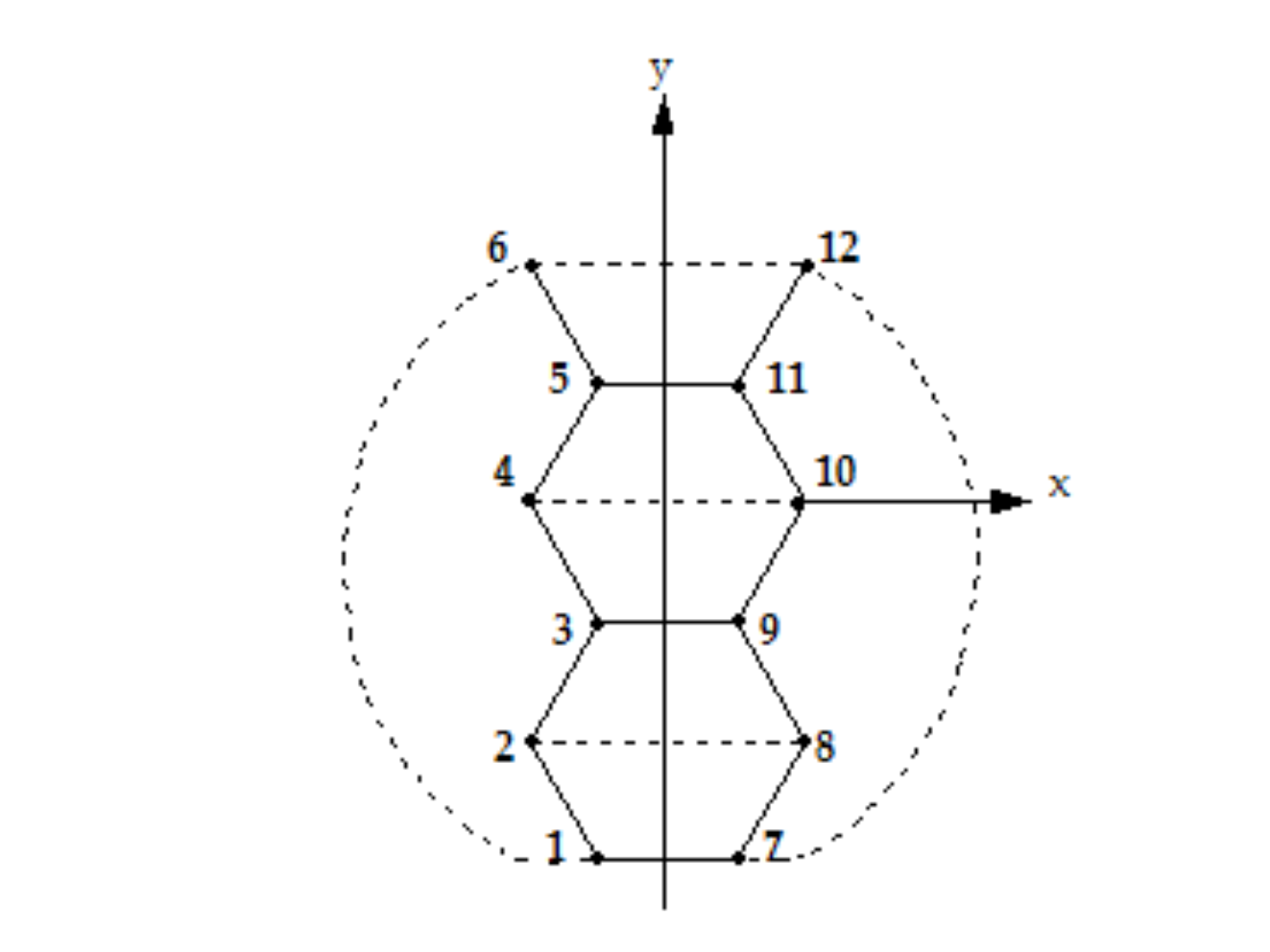}
\caption{Cluster $N=12$ for calculation of graphene wave functions.}
\label{Fig5}
\end{figure}

As an example, let us consider the cluster $N=12$, which is shown in Fig.\,\ref{Fig2b}, but with a different atoms numbering (Fig.\,\ref{Fig5}).

To determine the coefficients $c_1, c_2, \ldots, c_{12}$ in \eqref{eq:2} we must solve a system of twelve linear equations.
For cluster in Fig.\ref{Fig5} it has the form

%\begingroup
%\squeezetable
%\begin{table*}[htbp]

%\caption{}
%\begin{tabular}{rlllllllllllllr}
\begin{equation}
\begin{cases}

(1) \qquad \varepsilon c_1 + c_2 + c_6 + c_7 = 0 \\
(2) \qquad c_1 + \varepsilon c_2 + c_3 + c_8 = 0\\ 
(3) \qquad c_2 + \varepsilon c_3 + c_4 + c_9 = 0 \\ 
(4) \qquad c_3 + \varepsilon c_4 + c_5 + c_{10}= 0 \\
(5) \qquad c_4 + \varepsilon c_5 + c_6 + c_{11} = 0 \\ 
(6) \qquad c_1 + c_5 + \varepsilon c_6 + c_{12} = 0 \\ 
(7) \qquad c_1 + \varepsilon c_7 + c_8 + c_{12} = 0 \\ 
(8) \qquad c_2 + c_7 + \varepsilon c_8 + c_9 = 0 \\ 
(9) \qquad c_3 + c_8 + \varepsilon c_9 + c_{10} = 0 \\ 
(10) \qquad c_4 + c_9 + \varepsilon c_{10} + c_{11} = 0 \\ 
(11) \qquad c_5 + c_{10} + \varepsilon c_{11} + c_{12} = 0 \\ 
(12) \qquad c_6 + c_7 + c_{11}+\varepsilon c_{12} = 0 \\ 

\end{cases}
\label{eq:10}
%\end{aligned}
\end{equation}

%\end{table*}
%\endgroup

As we can see from Fig.\,\ref{Fig5} our cluster is reflection-symmetric in the axes $y$ and $x$, due to the closing bonds 1-6 and 7-12. Hence, wave functions must be symmetric ($S_x, S_y$) or antisymmetric ($A_{x}A_y$) to reflection in the $x$ and y axis. All together there are four possible symmetries of wave functions: $S_{x}S_y, S_{x}A_y, A_{x}S_y$ and $A_{x}A_y$.

First we consider the $S_{x}S_y$ case. Following relations among the coefficients must be satisfied

\begin{equation}
S_{x}S_y:
  \begin{cases}
c_1 = c_7\\
c_2 = c_6 = c_8 = c_{12}\\
c_3 = c_5 = c_9 = c_{11}\\
c_4 = c_{10}.
  \end{cases}
  \label{eq:11}
\end{equation}

Then the task is to solve just four equations instead of twelve.

If we take \eqref{eq:11}, the equations have the form

\begin{equation}
S_{x}S_y:
  \begin{cases}
1. (\varepsilon+1)c_1 + 2c_2 & = 0\\
2. \quad c_1 + (\varepsilon+1)c_2 + c_3 & = 0\\
3. \qquad c_2 + (\varepsilon+1)c_3 + c_4 & = 0\\
4. \qquad \qquad 2c_3 + (\varepsilon+1) c_4 & = 0.
  \end{cases}
  \label{eq:12}
\end{equation}

The determinant of \eqref{eq:12} has to be zero and therefore we obtain

\begin{equation}
S_x S_y: \varepsilon = -3, -2,0, 1.
\end{equation}

In other symmetry cases we have

\begin{equation}
S_{x}A_y:
  \begin{cases}
c_1 = -c_7,\\
c_2 = c_6 =-c_8= -c_{12},\\
c_3 = c_5 =-c_9= -c_{11},\\
c_4=-c_{10}.
  \end{cases}
  \label{eq:14}
\end{equation}

\begin{equation}
A_{x}S_y:
  \begin{cases}
c_1 = c_7 =0,\\
c_2 = - c_6 =c_8= -c_{12},\\
c_3 = -c_5 =c_9= -c_{11},\\
c_4=c_{10}=0.
  \end{cases}
  \label{eq:15}
\end{equation}

\begin{equation}
A_{x}A_y:
  \begin{cases}
c_1 = c_7 =0,\\
c_2 = - c_6 =-c_8=c_{12},\\
c_3 = -c_5 =-c_9= c_{11},\\
c_4=c_{10}=0.
  \end{cases}
  \label{eq:16}
\end{equation}

In the latter two cases there is $c_1=c_4=c_7=c_{10}=0$ due to the antisymmetry to axis $x$ ($A_x$). Therefore, only two equations remain in each case, namely those of the coefficients $c_2$ and $c_3$, and consequently only two values of energy for each $A_x S_y$ and $A_x A_y$.

The energy levels for various symmetries are shown in Tab.\,\ref{table2}, Tab.\,\ref{table3} shows the values of the coefficients $c_1, c_2, \ldots, c_{12}$. These coefficient values are obtained by solving the system of equations like \eqref{eq:12} and analogous systems for other symmetries for the energies $\varepsilon = 0. \pm 1, \pm 3$. The set of these coefficients determines the wave functions for various symmetries within our considered model. Tab.\,\ref{table2} and \ref{table3} also show the Brillouin zone symmetry points, which can also be seen in  Fig.\,\ref{Fig3}.

\begin{figure}[hptb!]
\centering
\subfigure[\label{Fig6a} $S_x S_y$; $K_{1}$]{\includegraphics[width=3.4cm]{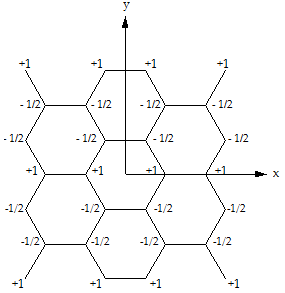}}
\qquad 
\centering
\subfigure[\label{Fig6b} $A_x S_y$; $K_{1}$]{\includegraphics[width=3.4cm]{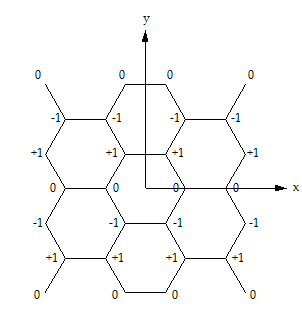}}
\subfigure[\label{Fig6c} $S_x A_y$; $K_{2}$]{\includegraphics[width=3.6cm]{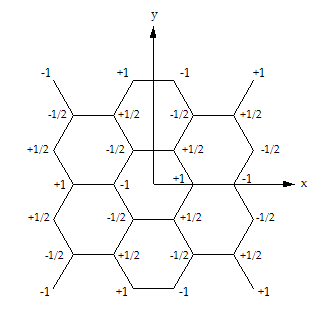}}\qquad
\subfigure[\label{Fig6d} $A_x A_y$; $K_{2}$]{\includegraphics[width=3.8cm]{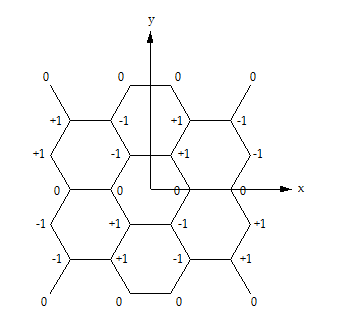}}
\caption{Wave functions of graphene various symmetries at the points $K$. The data is taken from Table\,\ref{table3}.}
\label{Fig6}
\end{figure}

For the purpose of illustration one can represent these functions graphically by extending the cluster over the whole graphene layer. In the case $\varepsilon = \pm 1$ the functions coincide with those in \cite{Wallace}. For the case $\varepsilon = 0$, wave functions for various symmetries are shown in Fig.\,\ref{Fig6}.

\begin{table*}
\caption{Energy levels of cluster $N=12$ corresponding to various symmetries of wave functions}
\begin{tabular}
 {>{\centering\arraybackslash}m{3.5 cm}|>{\centering\arraybackslash}m{1.0 cm}>{\centering\arraybackslash}m{1.0 cm}>{\centering\arraybackslash}m{1.0 cm}>{\centering\arraybackslash}m{1.0cm}>{\centering\arraybackslash}m{1.0cm}>{\centering\arraybackslash}m{1.0cm}>{\centering\arraybackslash}m{1.0cm}>{\centering\arraybackslash}m{1.0cm}>{\centering\arraybackslash}m{1.0cm}>{\centering\arraybackslash}m{1.0cm}>{\centering\arraybackslash}m{1.0cm}>{\centering\arraybackslash}m{1.0cm}}
\toprule[0.2 mm]
\hline
\multicolumn{1}{c|}{\backslashbox [35 mm] {\small Symmety} {\small Energy level}} & $\varepsilon_1$ & $\varepsilon_2$ & $\varepsilon_3$ & $\varepsilon_4$ & $\varepsilon_5$ & $\varepsilon_6$ & $\varepsilon_7$ & $\varepsilon_8$ & $\varepsilon_9$ & $\varepsilon_{10}$ & $\varepsilon_{11}$ & $\varepsilon_{12}$ \\ \midrule[0.2 mm]\hline
$S_x S_y$ & \multicolumn{1}{c}{-3} & \multicolumn{1}{c}{-2} &  &  & \multicolumn{1}{c}{0} &  &  &  & \multicolumn{1}{c}{1} &  &  &  \\ \hline
$S_x Y_y$ &  &  &  & \multicolumn{1}{c}{-1} &  & \multicolumn{1}{c}{0} &  &  &  &  & \multicolumn{1}{c}{2} & \multicolumn{1}{c}{3} \\ \hline
$A_x S_y$ &  &  & \multicolumn{1}{c}{-2} &  &  &  & \multicolumn{1}{c}{0} &  &  &  &  &  \\ \hline
$A_x A_y$ &  &  &  &  &  &  &  & \multicolumn{1}{c}{0} &  & \multicolumn{1}{c}{2} &  &  \\ \hline \midrule[0.2 mm]
Point of BZ & $\Gamma$ &  &  & $M - 0$ & $K_1$ & $K_2$ & $K_1$ & $K_2$ & $M + 0$ &  &  & D \\
\hline
\bottomrule[0.2 mm] 
\end{tabular}
\label{table2}
\end{table*}

\begin{table*}[htbp]
\begin{center}
\caption{Wave functions in the symmetry points of the Brillouin zone}
\begin{tabular}{c | c c c c c c c c}
\hline
%\toprule[0.2 mm]
\multicolumn{1}{c|}{\backslashbox{$c_n$}{Energy}} & -3 & -1 &  \vphantom{-} 0 &  \vphantom{-} 0 & \vphantom{-}  0 &  \vphantom{-} 0 & \vphantom{-}  1 & \vphantom{-}  3 \\ \hline \midrule[0.2 mm]
$c_1$ &  \vphantom{-} 1 &  \vphantom{-} 1 &  \vphantom{-} 1 &  \vphantom{-} 1 &  \vphantom{-} 0 &  \vphantom{-} 0 &  \vphantom{-} 1 &  \vphantom{-} 1 \\ 
$c_2$ &  \vphantom{-} 1 &  \vphantom{-} 1 & -$\nicefrac {1}{2}$ & -$\nicefrac {1}{2}$  &  \vphantom{-} 1 &  \vphantom{-} 1 & -1 & -1 \\ 
$c_3$ &  \vphantom{-} 1 &  \vphantom{-} 1 & -$\nicefrac {1}{2}$ & -$\nicefrac {1}{2}$ & -1 &  \vphantom{-} 1 &  \vphantom{-} 1 &  \vphantom{-} 1 \\ 
$c_4$ &  \vphantom{-} 1 &  \vphantom{-} 1 &  \vphantom{-} 1 & -1 &  \vphantom{-} 0 &  \vphantom{-} 0 & -1 & -1 \\ 
$c_5$ &  \vphantom{-} 1 &  \vphantom{-} 1 & -$\nicefrac {1}{2}$  & -$\nicefrac {1}{2}$  &  \vphantom{-} 1 & -1 &  \vphantom{-} 1 &  \vphantom{-} 1 \\ 
$c_6$ &  \vphantom{-} 1 &  \vphantom{-} 1 & -$\nicefrac {1}{2}$ & $\nicefrac {1}{2}$ & -1 & -1 & -1 & -1 \\ 
$c_7$ &  \vphantom{-} 1 & -1 & 1 & -1 &  \vphantom{-} 0 &  \vphantom{-} 0 &  \vphantom{-} 1 & -1 \\ 
$c_8$ &  \vphantom{-} 1 & -1 & -$\nicefrac {1}{2}$ & $\nicefrac {1}{2}$ &  \vphantom{-} 1 & -1 & -1 &  \vphantom{-} 1 \\ 
$c_9$ &  \vphantom{-} 1 & -1 & -$\nicefrac {1}{2}$ & $\nicefrac {1}{2}$ & -1 & -1 &  \vphantom{-} 1 & -1 \\ 
$c_{10}$ &  \vphantom{-} 1 & -1 &  \vphantom{-} 1 &  \vphantom{-} 1 &  \vphantom{-} 0 &  \vphantom{-} 0 & -1 &  \vphantom{-} 1 \\ 
$c_{11}$ &  \vphantom{-} 1 & -1 & -$\nicefrac {1}{2}$ & $\nicefrac {1}{2}$ &  \vphantom{-} 1 &  \vphantom{-} 1 &  \vphantom{-} 1 & -1 \\ 
$c_{12}$ &  \vphantom{-} 1 & -1 & -$\nicefrac {1}{2}$ & $\nicefrac {1}{2}$ & -1 &  \vphantom{-} 1 & -1 &  \vphantom{-} 1 \\ \hline
Symmetry & $S_x S_y$ & $S_x A_y$ & $S_{x} S_{y}$ & $S_x A_y$ & $A_x S_y$ & $A_x A_y$ & $S_x S_y$ & $S_x A_y$ \\ 
Point of BZ & $\Gamma$ & $M-0$ & $K_{1}$ & $K_{2}$ & $K_{1}$ & $K_{2}$ & $M+0$ & $D$ \\ \bottomrule[0.2 mm]\hline
\end{tabular}
\label{table3}
\end{center}
\end{table*}

From Fig.\,\ref{Fig6} it is easy to see, that in cases \ref{Fig6a} and \ref{Fig6b} the wave functions are plane waves with $\lambda = \frac{3}{2} a$, where $a$ is the length of the vector connecting the nearest neighbors of atoms A and B. The fronts of these waves are parallel to axis $x$, which means that the wave vector \textbf{\emph{k}} is directed along the axis $y$.

The length of the wave vector of these waves $k = \frac{2\pi}{\lambda} = \frac{4\pi}{3a}$ coincides with the length of \textbf{\emph{k}} in Brillouin zone point $K_1$ (Fig.\,\ref{Fig3}) - thus, Figures \ref{Fig6a} and \ref{Fig6b} show the wave functions at point $K_1$. In cases \ref{Fig6c} and \ref{Fig6d} one can easily see, that the fronts of the wave of each sublattice are at an $120^\circ$ angle to the $x$ axis. Therefore, the waves vectors are directed at a $30^\circ$ angle to the $x$ axis, that is the point $K_2$ in Fig.\,\ref{Fig3}. The length of these waves is also $\frac{3}{2} a$. As a consequence, Fig\,\ref{Fig6c} and Fig.\,\ref{Fig6d} show the wave functions at point $K_2$.

The fact that all four wave functions shown in Fig.\,\ref{Fig6} correspond to the energy $\varepsilon = 0$ can be directly deduced from the form of these functions. Namely, the total energy of the interactions of the atoms with their respective nearest neighbors is equal to zero, taking into consideration the values and signs of $c_n$ from Fig.\,\ref{Fig6}.

In Fig.\,\ref{Fig6} it is also apparent, that those structure elements, which are periodically repeated within the wave functions at $\varepsilon = 0$ are triplets of atoms: $1,-1,0$ (Fig.\,\ref{Fig6b}), $1, -\frac{1}{2}, -\frac{1}{2}$ (Fig.\,\ref{Fig6a}) etc. Consequently, the energy value $\varepsilon = 0$ is only possible in clusters, in which the number of atoms $N$ is divisible by three. Since $N$ also has to be even, due to the equal number of atoms in sublattice A and B, the number $N$ finally has to be divisible by six. This is the base for rule 3 of cluster building from section\,\ref{sec:essence}.

It is easy to see that in clusters $N=8,10$ and other $N$ not divisible to six, energy values $\varepsilon = 0$ are absent. Therefore, only clusters with by six divisible $N$ must be taken into consideration in the case of endless layers.

\section{Nanoribbons and nanotubes}
\label{sec:Nanoribbons and nanotubes}

\begin{figure} 
  \centering 
   \subfigure[\label{Fig7a}]{\includegraphics[width=4.5cm, height=2cm]{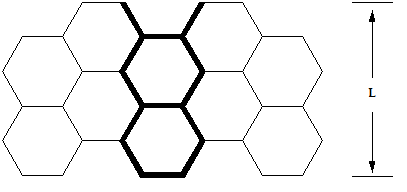}}\qquad 
   \subfigure[\label{Fig7b} Related closed cluster N=12]{\includegraphics[width=2cm, height=3.5cm]{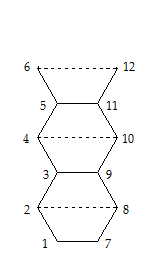}}
  \caption{Armchair ribbon (a) and the corresponding closed cluster $N=12$ (b).} 
  \label{Fig7}
\end{figure}

The CC approach is especially applicable to impurities, as well as nanotubes and nanoribbons. In the latter two cases a group of atoms lying across the tubes or ribbons is a natural cluster, in which the number of atoms depends on the diameter of the tube respectively the width of the ribbon (Fig.\,\ref{Fig7}). Let us now consider the armchair ribbon with $N=12$. As zigzag nanoribbons are not analogous to carbon nanotubes \cite{Wakabayashi2} the former will not be discussed in this work. The above mentioned cluster is shown in (Fig.\,\ref{Fig7b}). This cluster differs from the cluster in Fig.\,\ref{Fig2b} in a single, yet very significant aspect: the cluster in Fig.\,\ref{Fig7b} has no closing bonds between atoms 1-6 and 7-12. This reflects the fact, that the latter atoms are situated on the edge and therefore are connected to the ribbon by means of two, not three bonds.

As one can see, the cluster in Fig.\,\ref{Fig7b} belongs to a well known ladder type. The determinant $D_N$ of this cluster may be written in block form for any even $N$ 

\begin{equation} \label{eq:17}
D_N = \left| \begin{matrix} L_{N_1} (\varepsilon)&I_{N_1}\\ I_{N_1}&L_{N_1} (\varepsilon) \end{matrix} \right|,
\end{equation}

where 

\begin{equation*}
N_1 = \frac {N}{2}, 
\end{equation*}

\begin{equation} \label{eq:18}
\begin{aligned}
L_{N_1} (\varepsilon) = \left.\begin{array}{l}\left| \begin{matrix} \varepsilon& 1 &&&&&&\\ 
1&\varepsilon&1&&&&& \\
&1&\varepsilon&1&&&& \\
&&&\bold{.}&&&& \\
&&&&\bold{.}&&& \\
&&&&&\bold{.}&& \\
&&&&&1&\varepsilon&1 \\
&&&&&&1&\varepsilon
 \end{matrix} \right|\end{array}\right\}
 \end{aligned}
  N_1
\end{equation} \\

and $I_{N_1}$ being the unity matrix of order $N_1$. \\

The determinant \eqref{eq:17} can be brought into a quasi-triangular form. For this, we need to add row number $(N_1+1)$ to the first row, row $(N_1+2)$ to the second and so on until row $N_1$ is added to row $2N_1$. Then, the first column is subtracted from column $(N_1+1)$, second from $(N_1+2)$ and analogously column $N_1$ from $2N_1$. \\

As result we obtain

\begin{equation} \label{eq:19}
D_N = L_{N_1} (\varepsilon + 1) L_{N_1} (\varepsilon - 1),
\end{equation}

where $L_n(x)$ are well-known polynomials, expressions of which are given in [10] (for $g_n(x)$ there) for $n \leq 20$. It is also known that representation of $L_n(x)$ via the trigonometrical functions is [11]

\begin{equation} \label{eq:20}
L_n (x) = \frac{sin ((n+1) \theta)}{sin \theta},
\end{equation}

where $x=2cos \theta$.

Then, for obtaining the energy values $D_N (\varepsilon) = 0$ is reduced to conditions, which need to satisfy at least one of two of the following equations:

\begin{equation} \label{eq:21}
L_{N_1} (\varepsilon + 1) = 0
\end{equation} 

or

\begin{equation} \label{eq:22}
L_{N_1} (\varepsilon - 1) = 0.
\end{equation}

In view of \eqref{eq:20}, these equations lead to

\begin{equation} \label{eq:23}
sin[(N_1 + 1) \theta_+] = 0
\end{equation}

and

\begin{equation} \label{eq:24}
sin[(N_1 + 1) \theta_-] = 0,
\end{equation}

where the $\theta_{\pm}$ are determined from

\begin{equation} \label{eq:25}
2cos \theta_{\pm} = \varepsilon \pm 1 .
\end{equation}

The solutions of \eqref{eq:23} and \eqref{eq:24} are 
\begin{equation} \label{eq:26}
\theta_{\pm} = \frac{n_\pm \pi}{N_1 + 1}, 
\end{equation}

where $n_{\pm}$ = 1,2,\ldots,$N_1$.

Then, from \eqref{eq:25} we get the formula for the energy values of a closed cluster with $2N_1$ atoms, which describes the nanoribbon

\begin{equation} \label{eq:27}
\varepsilon_{n_\pm} = 2cos \frac{n_\pm \pi}{N_1 + 1} \pm 1 .
\end{equation}

Taking into account that $n_+$ and $n_-$ both may possess $N_1$ values and due to the $\pm 1$ part we see from \eqref{eq:27}, that $\varepsilon_{n_\pm}$ possesses $2N_1=N$ values, as this has to be the case for a cluster with $N$ atoms.

If the ribbon is transformed to an infinite layer $(N_1 \rightarrow \infty)$, then we get the boundary energy spectrum values from \eqref{eq:27}, which are $\varepsilon_{max} = 3$ at $n_+=1$ and $\varepsilon_{min} = - 3$ at $n_-=N_1$. These values also coincide with the result in section\,\ref{sec:Ham}.

Using equation \eqref{eq:27} one can easily find the values of $N_1$, under which the solutions $\varepsilon_{n \pm}=0$ exist, that is the ribbon is metallic. The condition for this is $cos \frac{n_-\pi}{N_1+1} = \frac{1}{2}$ or $cos \frac{n_+ \pi}{N_1+1} = -\frac{1}{2}$, otherwise $\frac{n_- \pi}{N_1+1} = \frac{\pi}{3}$, $\frac{n_+ \pi}{N_1+1} = \frac{2\pi}{3}$; $n_+ = 2n$. Denoting $n_{-} = M$ one can write the both latter equations as

\begin{equation} \label{eq:28}
N_1 = 3M - 1,
\end{equation}
with M being an integer.

The formula \eqref{eq:28} is the same as has been obtained by means of the tight binding model in \cite{Fujita} and \cite{Nakada}.

The wave functions of nanoribbons with metallic conductivity can be found by using the symmetry like it is done for the infinite layer in section\,\ref{sec:Wave functions}.

\begin{figure}[H] 
  \centering 
   \subfigure[\label{Fig8a}]{\includegraphics[width=1.7cm]{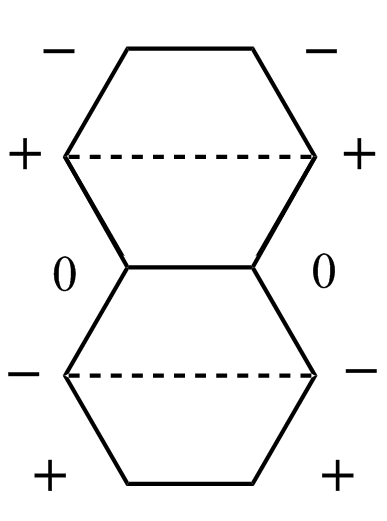}}\
   \subfigure[\label{Fig8b}]{\includegraphics[width=1.7cm]{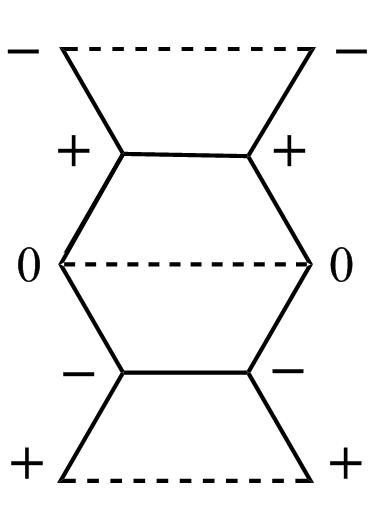}}\ 
   \subfigure[\label{Fig8c}]{\includegraphics[width=1.8cm]{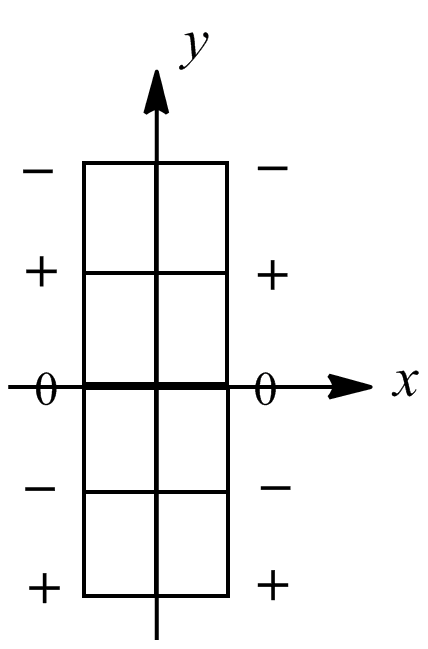}}\ 
   \subfigure[\label{Fig8d}]{\includegraphics[width=3cm]{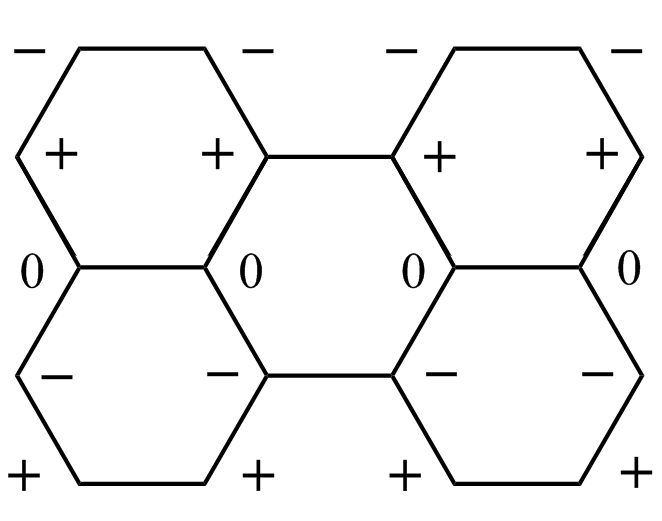}}
  \caption{Wave functions of clusters for description of ribbon $N_1=5$ (a,b), ladder cluster (c) and ribbon with symmetry $A_x S_y$ at $\varepsilon=0$ (d).} 
  \label{Fig8}
\end{figure}

We consider as an example the case $N_1= 5$. The closed clusters for the description of the ribbon are shown in figures \ref{Fig8a} and \ref{Fig8b}. According to the graph theory terminology \cite{Trinajstic}, these clusters are isomorphic to each other as well as to the ladder graph shown in Fig.\,\ref{Fig8c}. The energy spectra of these graphs are therefore the same.

Likewise to the case of infinite layer one can also obtain the energy values of the cluster, which are corresponding to various symmetries. Results for $N_1=5$ are shown in Tab.\,\ref{table4}.

\begin{table*}
\caption{Energy spectrum of a closed cluster for description of a $N_1=5$ ribbon}
\begin{tabular} {>{\centering\arraybackslash}m{1.2 cm}|>{\centering\arraybackslash}m{1.2 cm}|>{\centering\arraybackslash}m{1.2cm}|>{\centering\arraybackslash}m{1.2cm}|>{\centering\arraybackslash}m{1.2cm}|>{\centering\arraybackslash}m{1.2cm}|>{\centering\arraybackslash}m{1.2cm}|>{\centering\arraybackslash}m{1.2cm}|>{\centering\arraybackslash}m{1.2cm}|>{\centering\arraybackslash}m{1.2cm}|>{\centering\arraybackslash}m{1.2cm}}

\toprule[0.1 mm] \hline \tiny {Symmetry} &  \multicolumn{10}{c}{\tiny{Energy}}   \\
\midrule[0.1 mm] \hline {$S_x S_y$} & {$-\sqrt{3} -1$ }& \hspace*{0.3 cm} & {-1} & \hspace*{0.3 cm} & \hspace*{0.3 cm} & \hspace*{0.3 cm} & {$\sqrt{3} -1$ }& \hspace*{0.3 cm} & \hspace*{0.3 cm} & \hspace*{0.3 cm} \\ 
\hline {$S_x A_y$} & \hspace*{0.3 cm} & \hspace*{0.3 cm} & \hspace*{0.3 cm} & {$-\sqrt{3} + 1$ } & \hspace*{0.3 cm} & \hspace*{0.3 cm} & \hspace*{0.3 cm} & {1} & \hspace*{0.7 cm} & { $\sqrt{3} + 1$ }\\ 
\hline {$A_x S_y$} & \hspace*{0.3 cm} & {-2} & \hspace*{0.3 cm} & \hspace*{0.3 cm} & {0} & \hspace*{0.3 cm} & \hspace*{0.3 cm} & \hspace*{0.3 cm} & \hspace*{0.3 cm} & \hspace*{0.3 cm} \\ 
\hline {$A_x A_y$} & \hspace*{0.3 cm} & \hspace*{0.3 cm} & \hspace*{0.3 cm} & \hspace*{0.3 cm} & \hspace*{0.3 cm} & {0} & \hspace*{0.3 cm} & \hspace*{0.3 cm} & {2} & \hspace*{0.3 cm} \\
\hline 
\end{tabular} 
\label{table4} 
\end{table*}

If only the energy spectrum is required it can be directly obtained from equation \eqref{eq:27}.

\begin{figure}[htbp]
\centering
\includegraphics[width=4.2cm, height=4cm]{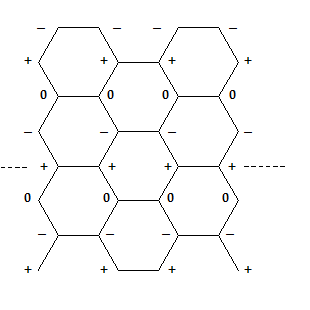}
\caption{Wave functions of symmetry $A_x S_y$ for the ribbon $N_1=8$ at $\varepsilon=0$.}
\label{Fig9}
\end{figure}

The cases for other $N_1$ can be considered analogously. Figure \ref{Fig9} shows the wave functions of the symmetry $A_x S_y$ for the $N_1=8$ ribbon at $\varepsilon = 0$.

If we compare the ribbons shown in Figures \ref{Fig8d} and \ref{Fig9} we can see the meaning of the metallic conductivity condition \eqref{eq:28}: when increasing $N_1$, in order to keep the energies of each atom at zero, it is necessary to add two rows of atoms, with positive and negative $\ket{p_z}$. The latter must however be separated from the previous row by including a row of zeroes. Altogether, we need to add three rows, which is where the term $3M$ in \eqref{eq:28} derives from.
From Figures \ref{Fig8} and \ref{Fig9} one can also see that the distance between the nodes $\frac{\lambda}{2}=\frac{3}{4} a$ and the wave front is parallel to axis $x$. This corresponds to the wave vector ${\boldsymbol{k}}$  at point $K_1$ of the Brillouin zone. It is easy to see, that in case of the symmetry $A_x A_y$ the energy value $\varepsilon = 0$ is corresponding to the point $K_2$, similar to the infinite layer.

Using $N_1$ is not the only way to describe the criterion for metallic conductivity. It also can be written by the means of the width of the ribbon $L$. Figures \ref{Fig8d}, \ref{Fig9} and equation \eqref{eq:28} show that it is possible to obtain the connection between $M$, $N_1$ and $L$ as in Table\,\ref{table5}.

\begin{table}
\caption{The connection between the ribbon width $L$ (scaled by a), $N_1$ and $M$ in the case of metallic conductivity.}
\begin{tabular}{|c|c|c|}
\toprule[0.2 mm]
\hline $M$ & $N_1$ & $L^{met}$ \\
\midrule[0.2 mm]
\hline 1 & 2 & $\nicefrac{1}{2}$ \\ 
\hline 2 & 5 & 2 \\ 
\hline 3 & 8 & $\nicefrac{7}{2}$ \\ 
\hline 4 & 11 & 5 \\ 
\hline \ldots & \ldots & \ldots \\
\hline
\end{tabular}

\label{table5}
\end{table}

The relation between $L$ and $M$ in Table\,\ref{table5} can be written as

\begin{equation} \label{eq:29}
L^{met} = \frac{3M - 2}{2}, \qquad  \textnormal{$M = 1,2,3$ \ldots} \quad.
\end{equation}

This formula differs from  $L = 3M +1$ in \cite{Brey}, which is maybe due to the different definition of the ribbon width in \cite{Brey}.

\begin{figure}[htbp]
\includegraphics{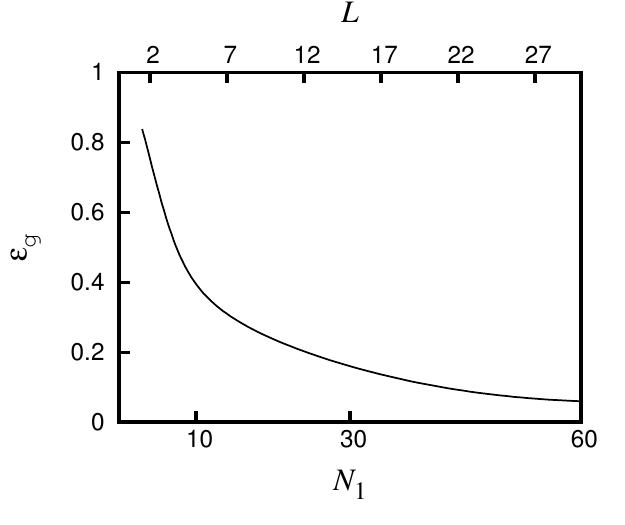}
\caption{Dependence between the energy gap $\varepsilon_g$ and the ribbon width.}
\label{Fig10}
\end{figure}

Figure \eqref{Fig10} shows the dependence between the energy gap $\varepsilon_g$ and the ribbon width. Here it is $\varepsilon = 2\varepsilon_1$, where $\varepsilon_1$ is the nearest level to zero and is found via \eqref{eq:27}.

The values of $N_1$ and L, where $\varepsilon_g = 0$ are not marked, they can be obtained from equations \eqref{eq:28} and \eqref{eq:29}.

This described approach of the calculation of nanoribbons by the means of ladder clusters may also be applied to nanotubes, with some modifications. The formal transition from the nanoribbon to the nanotube consists in closing a cluster of the type shown in Fig.\,\ref{Fig7b}, so that closing happens between atoms 1-6 and 7-12 - this way we get the cluster from, Fig.\,\ref{Fig2b}. If we wrap it up into a cylinder form, we obtain a cluster with three hexagons. It can be easily verified that in clusters of this type the number of hexagons in the cylinder section is always an integer.

However, a cluster of type as in Fig.\,\ref{Fig8} cannot be wrapped up in such a way, because the atoms of the upper and lower edge are from the same sublattice, A or B - and nearest neighbors always have to be from different sublattices.

Therefore, in order to wrap up a cluster into a cylinder one must add a row of atoms. These atoms need to belong to a different sublattice than the atoms at the edges of the initial cluster. In an $N_1=5$ example this means that the number $N_c$ of atoms situated along the circle cylinder section must equal six.

It may also be easily verified that for clusters of both types $N_c$ must be even (which is a consequence of the equal number of A and B atoms) and the number of hexagons $N_S$ along the circle has to be an integer. Hereat, beginning with the smallest value $N_s=2$ it is $N_c=2N_s$.

In case of metallic conductivity ($\varepsilon_g =0$) as well as in the infinite layer (section\,\ref{sec:Wave functions}), it is necessary for $N_c$ to contain an integer number of atomic triplets $+,-,0$, that is, $N_c$ needs to be not only even but also divisible by three. All in all $N_c $ has to be divisible by six, and therefore we can write the criterion of metallic conductivity of nanotubes as 

\begin{equation}
N_c^{met} = 6 M, \qquad M=1,2,3\ldots.
\label{eq:30}
\end{equation}

For all other values of $N_c$ there is $\varepsilon \neq 0$. For example, in case of the least values $N_s=2, N_c=4$ one must consider the closed cluster $N=8$. It is possible to calculate a cluster of type from Figure \ref{Fig2b}, but for $N=8$.

Solving equation \eqref{eq:4} for this case gives us the following energy spectrum:

\begin{equation}
\varepsilon^c = \pm 1, \pm 1, \pm 1, \pm 3.
\label{eq:31}
\end{equation}

Hence, the value of the energy gap for $N_c=4$ is $\varepsilon^c_g = 2$, which is the maximum value of $\varepsilon^c_g$ for nanotubes. With an increasing $N_c$ the value of $\varepsilon^c_g$ decreases.

It should be noted, that $\varepsilon^c$ in \eqref{eq:31} and $\varepsilon^c_g$ are scaled by $\gamma_0^c$: $\varepsilon_c = \frac{E_c}{\gamma_0^c}$, where $\gamma_0^c$ is the tube's transfer integral (hopping energy). This latter value is different from the value of $\gamma_0$ for the infinite layer and is dependent from $N_c$, which should be taken into the account when $\varepsilon _g ^c$ in $eV$ needs to be obtained. However, this dependence does not exert any influence on the criterion $\varepsilon_g = 0$.

\section{Impurity in graphene}
\label{sec:Impurity}

The cluster approach is, as already mentioned, especially appropriate for studying the impurities in crystals. That is connected with the fact that the impurity influence on the crystal depends mainly on the interaction with atoms, which are located in the nearest environment of this impurity. Exactly those atoms together with the impurity are the ones which can be considered as a cluster with N atoms. The study of this cluster is the essence of the impurity problem in the present approach.

In this paper we restrict the consideration to the simplest cluster $N=6$. Let us assume, that in the first lattice point of this cluster atom C is substituted by another one. This atom only differs from C in terms of the $\ket{p_z}$-electron energy of an isolated atom, which is $E_0 + \Delta E_0$ instead of just $E_0$. It is also assumed, that this impurity atom is silicon, although is also might be germanium or any other atom with the same outer electron shell structure as C. Furthermore, we assume the transfer integral $\gamma_0$ equal between all neighbor atoms. In order to account for the difference in these parameters within the cluster approach, one simply has to replace the ones in the corresponding Hamiltonian matrix elements with parameters, which characterize the transfer integral between the impurity and the surrounding atoms. The issue of the influence of these parameters on the energy spectrum in 3D crystals has been investigated by us earlier by means of Green's functions \cite{Tal2}. For the purpose of this paper, however, this problem is not of a fundamental meaning, in so far as the main parameter is $\Delta E_0$.

\begin{figure}[htbp]
\centering
\includegraphics[trim=0cm 0cm 0cm 1cm, width=2.3cm]{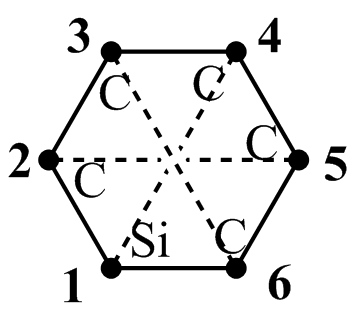}
\caption{Closed cluster $N=6$ with one impurity atom Si.}
\label{Fig11}
\end{figure}

An example of a closed cluster with one impurity atom Si is shown in Fig.\,\ref{Fig11}.

The secular equation for this cluster has the form

\begin{equation}
D_{6} =	\left|   
\begin{matrix}
(\varepsilon-\Delta) & 1 & \hspace*{0.7 cm} & 1 & \hspace*{0.7 cm} & 1 \\ 
1 & \varepsilon & 1 & \hspace*{0.7 cm} & 1 & \hspace*{0.7 cm} \\ 
\hspace*{0.7 cm} & 1 & \varepsilon & 1 & \hspace*{0.7 cm} & 1 \\ 
1 & \hspace*{0.7 cm} & 1 & \varepsilon & 1 & \hspace*{0.7 cm} \\ 
\hspace*{0.7 cm} & 1 & \hspace*{0.7 cm} & 1 & \varepsilon & 1 \\ 
1 & \hspace*{0.7 cm} & 1 & \hspace*{0.7 cm} & 1 & \varepsilon
\end{matrix} 
\right|=0,
\label{eq:32}
\end{equation}

where $\Delta = \frac{\Delta E_0}{\gamma_0}$.

This sixth degree equation is reduced to a cubic equation, which can be solved analytically. As result we get following solutions

\begin{equation}
\begin{aligned}
\varepsilon_{1,2,3} &= 0.\\
\varepsilon_{4,5,6} &= \frac{1}{3} \left[ \Delta + 2 \sqrt{\Delta^2 + 27} \ \cos \left( \frac{\varphi}{3} + n \frac{2 \pi}{3} \right)   \right],\\
&\textit{} \quad n=0,1,2\ldots
\end{aligned}
\label{eq:33}
\end{equation}

where

\begin{equation}
\cos \varphi = \frac{\Delta \left(\Delta^2 - \frac{3^4}{2} \right)}{\left( \Delta^2 + 27 \right)^{\frac{3}{2}}}.
\label{eq:34}
\end{equation}

It is easy to see that under $\Delta = 0$ there is

$$ \varepsilon^0_4 = 3, \varepsilon^0_5 = -3, \varepsilon^0_6 = 0, $$

that is we obtain the result \eqref{eq:6}. \\

\begin{figure}[htbp]
\centering
\includegraphics[trim=5cm 1.8cm 0cm 3cm, height=8cm]{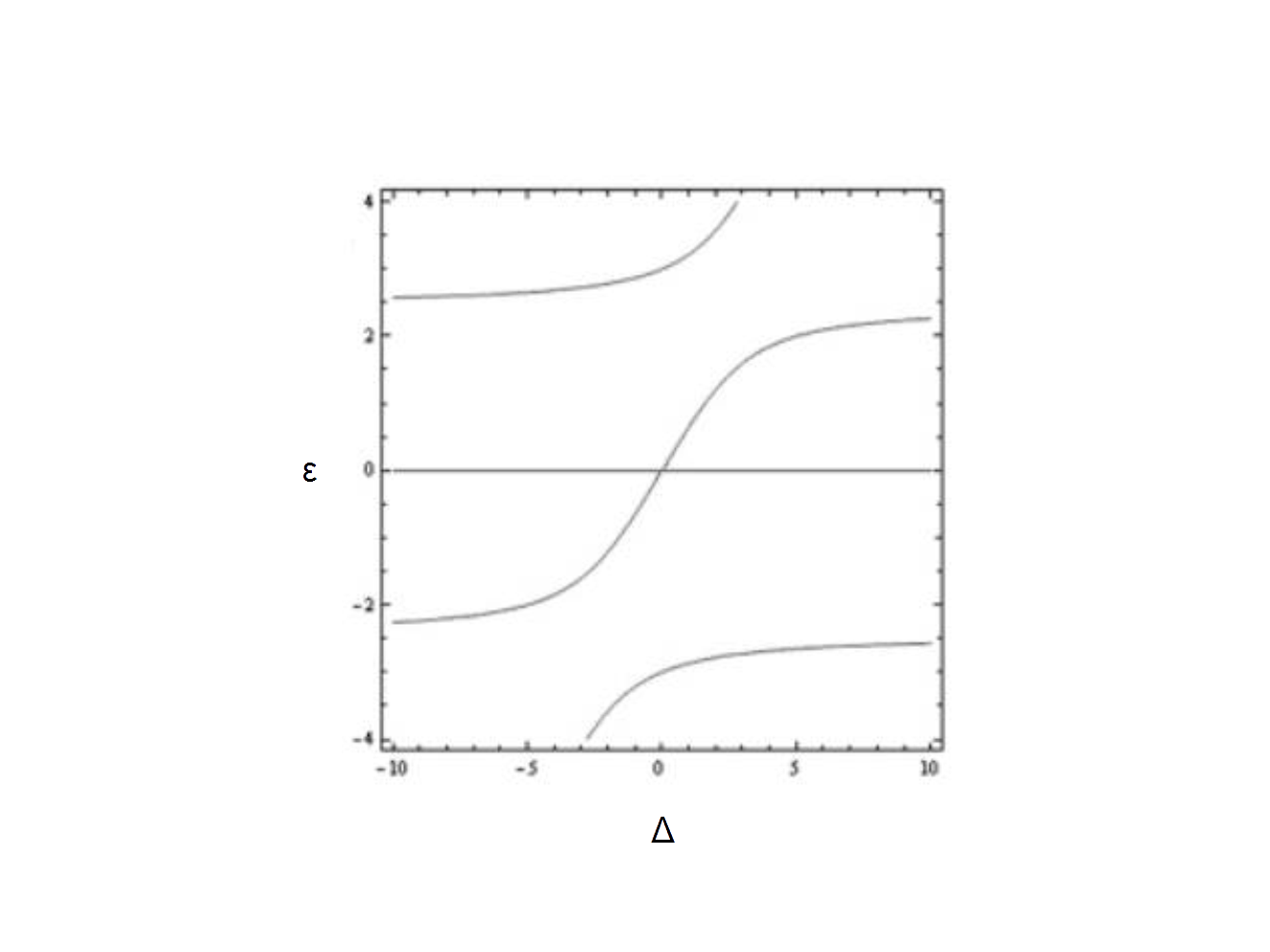}
\caption{Dependence of energy levels of cluster $N=6$ on $\Delta$. The level $\varepsilon = 0$ is threefold degenerated.}
\label{Fig12}
\end{figure}

The numerical calculations of the dependence of $\varepsilon$ from $\Delta$, which may be obtained from \eqref{eq:33} and \eqref{eq:34} or, simpler, directly from the calculation of the determinant \eqref{eq:32}, are shown in Figure \ref{Fig12}.

\qquad
As one can see, we get a result which is typical for the perturbation process: the fourfold degenerate level $\varepsilon=0$ splits partially and the non-degenerate level moves. At the same time level $\varepsilon=0$ still remains threefold degenerate. That is why there is no arising gap between the valence and conduction bands. The level, which is split from the originally fourfold degenerated level $\varepsilon=0$ is the impurity level in the conduction band, that is the resonance level.

\section{The energy bands of monoatomic layer 
C$_{1-x}$ Si$_{x}$ $(0\le x \le 1)$}
\label{sec:monoatomic layer}

By means of the CC approach one may as well calculate the energy spectrum of a hypothetical object, a monoatomic layer of the type C$_{1-x}$ Si$_{x}$. That such an object can be created results from the following reasoning: by obtaining by means of epitaxi, graphene can be formed on the surface of SiC. The monolayer of Si, which appear hereat as a buffer layer, has also a hexagonal structure \cite{Trauzettel}. That is why by means of epitaxi, perhaps not only graphene can be formed but also the monoatomic layer of the type C$_{1-x}$ Si$_{x}$ $(0\le x \le 1)$, with a hexagonal structure similar to graphene. 

In the study of this problem we restrict ourselves to consider only the simplest structure $N=6$. This is found to be sufficient for describing in outline the energy band structure of C$_{1-x}$ Si$_{x}$.

We consider the cases separately when in a cluster of 6 C-atoms one, two, three, four or five atoms are replaced by Si-atoms. This corresponds to the values $x=\frac{1}{6};\frac{1}{3};\frac{1}{2}; \frac{2}{3};\frac{5}{6}$, respectivaly. The case of $x=\frac{1}{6}$ has been already considered in the preceding section. It should be noted that under "impurity" we mean the atoms which are present in a lesser number. That is at $x=\frac{1}{3}$ the impurity atoms are those of Si and at $x=\frac{2}{3}$ those of C. 

\begin{figure}[H] 
  \centering 
   \subfigure[\label{Fig13a} $x=1/3$]{\includegraphics[width=2cm]{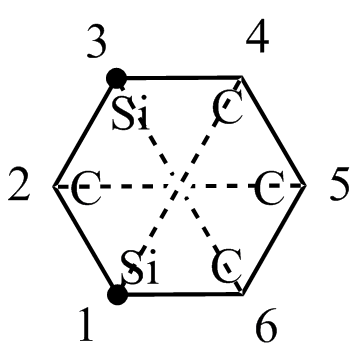}}\
   \subfigure[\label{Fig13b} $x=1/2$]{\includegraphics[width=2cm]{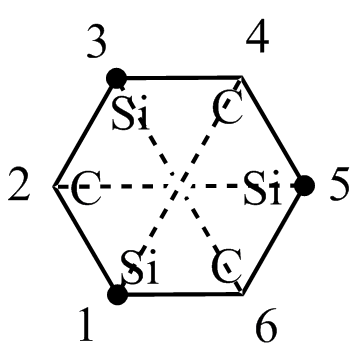}}\ 
   \subfigure[\label{Fig13c} x=2/3]{\includegraphics[width=2cm]{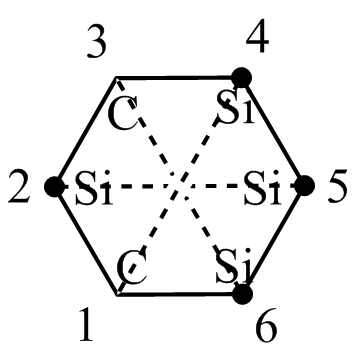}}\ 
   \subfigure[\label{Fig13d} $x=5/6$]{\includegraphics[width=2cm]{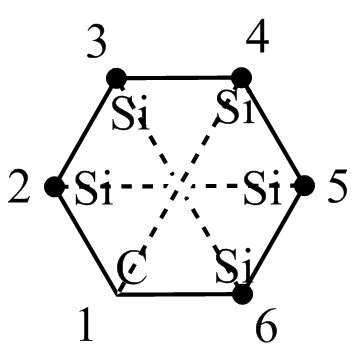}}
  \caption{Clusters of type $I$ with $N=6$ for C$_{1-x}$ Si$_{x}$.} 
  \label{Fig13}
\end{figure}

The clusters, corresponding to these $x$-values, can be of two types. Namely, by $x=\frac{1}{3}$ two impurity Si-atoms can interact with each other (type $II$) or not interact (type $I$). The same can be obtained for C-atoms by $x=\frac{2}{3}$. First we study the clusters of type $I$, which are shown in Fig.\,\ref{Fig13}. 

The secular equations for clusters shown in Fig.\,\ref{Fig13} can easily be estimated. For example, by $x=\frac{1}{3}$ we obtain

\begin{equation}
\begin{vmatrix}

(\varepsilon-\Delta) & 1 & \hspace*{0.7 cm} & 1 & \hspace*{0.7 cm} & 1 \\
1 & \varepsilon & 1 & \hspace*{0.7 cm} & 1 & \hspace*{0.7 cm} \\
\hspace*{0.7 cm} & 1 & (\varepsilon-\Delta) & 1 & \hspace*{0.7 cm} & 1 \\
1 & \hspace*{0.7 cm} & 1 & \varepsilon & 1 & \hspace*{0.7 cm} \\
\hspace*{0.7 cm} & 1 & \hspace*{0.7 cm} & 1 & \varepsilon & 1 \\
1 & \hspace*{0.7 cm} & 1 & \hspace*{0.7 cm} & 1 & \varepsilon \\

\end{vmatrix} =0.
\label{eq:35}
\end{equation}

For other values of $x$ the equations have the similar form. In all cases the equations of the sixth degree are to be reduced to equations of the third degree and are solved analytically.

The results of these calculations are: 

\begin{equation*} 
x=\frac{1}{3}:  \qquad \varepsilon_{1,2}=0;\ \varepsilon_{3}=\Delta 
\end{equation*}
 
$\varepsilon_{4,5,6}$ are determined by formula \eqref{eq:33}, but with other $\varphi$, namely

\begin{equation}
\cos \varphi=\frac{\Delta^{3}}{(\Delta^{2}+27)^\frac{3}{2}}.
\label{eq:36}
\end{equation}

\begin{equation}
\begin{aligned}
x=\frac{1}{2},\quad (\mbox{SiC}):\\
&\varepsilon_{1,2} = 0;\\
&\varepsilon_{3,4} = \Delta;\\
&\varepsilon_{5,6} = \frac{\Delta\pm\sqrt{\Delta^2 + 36}}{2}.
\end{aligned}
\label{eq:37}
\end{equation}

\begin{equation} 
\begin{aligned}
x=\frac{2}{3}: \qquad
&\varepsilon_{1} = 0;\\
&\varepsilon_{2,3} = \Delta;\\
&\varepsilon_{4,5,6} = \frac{2}{3} \left[ \Delta - \sqrt{\Delta^2 + 27} \ \cos \left( \frac{\varphi}{3} + n \frac{2 \pi}{3} \right)   \right];\\
& n=0,1,2\ldots
\end{aligned}
\label{eq:38}
\end{equation}

where $\varphi$ is the same as in \eqref{eq:36}

\begin{equation}
\begin{aligned}
x=\frac{5}{6}: \qquad
&\varepsilon_{1,2,3} = \Delta;\\
&\varepsilon_{4,5,6} = \frac{2}{3} \left[ \Delta + \sqrt{\Delta^2 + 27} \ \cos \left( \frac{\varphi}{3} + n \frac{2 \pi}{3} \right)   \right];\\
\text{where}\\
&
\cos\varphi=\frac{\Delta(\Delta^{2}-\frac{1}{2}3^4)}{(\Delta^{2}+27)^\frac{3}{2}};\\
&\textit{} n=0,1,2\ldots
\end{aligned}
\label{eq:39}
\end{equation}

\begin{figure}[htbp!]
\centering
\includegraphics[width=7cm, height=11.5cm]{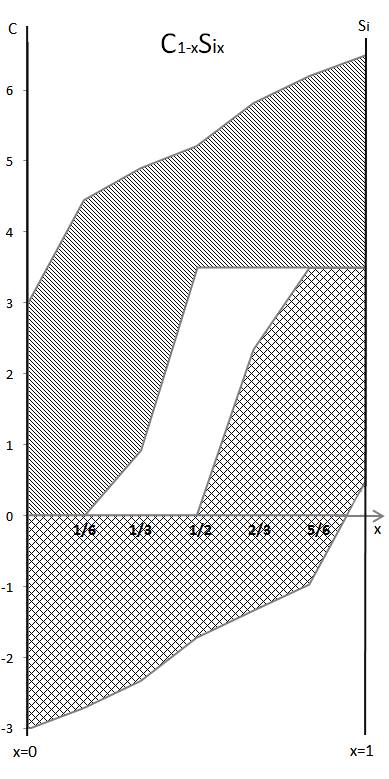}
\caption{Energy spectrum of the mono-atomic layer C$_{1-x}$ Si$_{x}$. checkered: valence band; striped: conduction band; plain: forbidden band}
\label{Fig14}
\end{figure}

The dependence of the energy bands of C$_{1-x}$ Si$_{x}$ on $x$ is shown in form of a diagram in Fig.\,\ref{Fig14}. This diagram is obtained from formulas \eqref{eq:33},\eqref{eq:34},\eqref{eq:36}-\eqref{eq:39} for $\Delta=3.5$ which corresponds roughly to the value of $\Delta$ for Si in graphene.

The main feature of this band structure rises from forbidden band with the typical tunnel form. From this one can suggest, that in principle it is possible to create a tunnel diode on the basis on a C$_{1-x}$ Si$_{x}$ layer. In the ground of such a diode can be a tunnelling between Si- and C-domains of the layer in which the concentration of $x$ changes with the coordinate $x$. That is, if in Fig\,\ref{Fig14} $x$ is not only the concentration, but as well the coordinate $x$.

Next we study the clusters of type $II$ in which impurity atoms interact not only with basic atoms of the layer but also with each other. The latter is possible, if the impurities are in the lattice points of different sublattice: A and B, because the atoms of the same sublattice do not interact with each other in the nearest neighbors approximation.

In case of $x=\frac{1}{3}$ the cluster of type $II$ can be obtained from the cluster in Fig.\,\ref{Fig13a}, if place the atoms Si in the lattice points, for example 2 and 5 instead of 1 nad 3. C-atoms with $x=\frac{2}{3}$ in the cluster in Fig.\,\ref{Fig13c} can be treated the same way. As a result we obtain following solutions for clusters of type $II$:

\begin{equation}
\begin{aligned}
x=\frac{1}{3}: \qquad
\varepsilon_{1,2} &= 0;\\
\varepsilon_{3,4} &= \frac{1}{2} \left( \Delta - 3 \pm \sqrt{\Delta^2 + 2\Delta + 9} \right);\\
\varepsilon_{5,6} &= \frac{1}{2} \left( \Delta + 3 \pm \sqrt{\Delta^2 - 2\Delta + 9} \right).\\
\end{aligned}
\label{eq:40}
\end{equation}

\begin{equation}
\begin{aligned}
x=\frac{2}{3}: \qquad
\varepsilon_{1,2} &= \Delta;\\
\varepsilon_{3,4} &= \frac{1}{2} \left( \Delta - 3 \pm \sqrt{\Delta^2 - 2\Delta + 9} \right);\\
\varepsilon_{5,6} &= \frac{1}{2} \left( \Delta + 3 \pm \sqrt{\Delta^2 + 2\Delta + 9} \right).\\
\end{aligned}
\label{eq:41}
\end{equation}

From formulas \ref{eq:40} and \ref{eq:41} one can obtain the energy of impurity states, which at $\Delta=3.5$ can be found most within valance and conduction bands. The single level which can be found in the forbidden band appears when $x=\frac{1}{3}$. This level can take part in tunnelling, which we discussed above.

\section{Conclusions}
\label{sec:Con}

In this paper, the CC approach has been applied to some problems of calculation of the grpahene energy spectrum. This approach seems to be especially appropriate both for problems of the breach of the graphene's periodical structure (point defects, boundaries) and for problems besides monolayer graphene (bilayer graphene, compound C$_{1-x}$ Si$_{x}$).

As for the precision of the approach, it may easily be increased by considering clusters of larger sizes and interaction not only with the nearest neighbors. Besides, it is not difficult to study the cluster with a distortion of the lattice near the point defect. One can also consider the conglomerates of point defects in a graphene lattice.

Except various problems with the calculation of electron energy spectrum, the closed cluster approach may as well be applied to calculate of the vibration spectrum in graphene, likewise as it is done for one-dimensional and three-dimensional crystals in \cite{Tal3}, \cite{Tal4}.

\section{Acknowledgments}
\label{sec:X}

I would like to thank Olga Talianska and Raisa Kociurzynski for the interest and informative support, Alexander Talyanski for help with the numerical calculations and Ludmila Swarytsch for the motivation of this work. Especially, I wish to thank Andr\'{e} and Raisa Kociurzynski for helping preparing the publication. 

\pagebreak

\end{document}